\documentclass[preprint,12pt]{elsarticle}

\usepackage{graphics}

\usepackage{amssymb}

\journal{Physics Letters A}

\begin{document}

\begin{frontmatter}



\title{Singularities in ion trap nonlinear coherent states}


\author[poli1]{F. A. Raffa\corref{cor1}}
\ead{francesco.raffa@polito.it}

\author[poli2,isifound]{M. Rasetti}

\author[rim]{M. Genovese}

\cortext[cor1]{Corresponding author}

\address[poli1]{Dipartimento di Meccanica, Politecnico di Torino \\Corso Duca degli Abruzzi, 24 - 10129 Torino, Italy}
\address[poli2]{Dipartimento di Fisica, Politecnico di Torino \\ Corso Duca degli Abruzzi, 24 - 10129 Torino, Italy}
\address[isifound]{ISI Foundation, Viale Settimio Severo, 35 - 10133 Torino, Italy}
\address[rim]{Divisione di Ottica, INRIM, Strada delle Cacce, 91 - 10135 Torino, Italy}

\begin{abstract}
We present a theoretical analysis of the $k$-boson nonlinear coherent states of a two-level trapped ion interacting with two laser fields. Such states are both the zero-energy state of the interaction Hamiltonian and the eigenstates of a deformed annihilation operator. For the single-boson case, we show that the structure of the states and their coherence and minimum-uncertainty properties can be compromised whenever the Lamb-Dicke parameter is one of the roots of certain Laguerre polynomials. We investigate these problems, which are strictly related to the non-analyticity of the deformation function in the annihilation operator.
\end{abstract}


\end{frontmatter}


\section{Introduction}
Nonlinear coherent states (NLCS) have become a tool of the utmost importance in quantum optics and many efforts have been devoted to investigating them, also with respect to their prospective applications in the growing field of quantum technologies (QT) such as quantum information (coding, transmission and elaboration of information by exploiting properties of quantum states), quantum computing, quantum imaging and quantum metrology.

Born from the extension of the Glauber states, various forms of NLCS have been constructed and analyzed from both the theoretical and the experimental point of view.

In the present paper the focus is on the $k$-boson and single-boson NLCS which can be realized by means of a trapped ion interacting with driving electromagnetic fields as reported in \cite{VOGEL1,MATOS1}. It is shown that for (and close to) certain values of the Lamb-Dicke parameter $\eta$, the original Vogel NLCS are strongly modified resulting into new different states, whose experimental realization appears problematic. The basic quantum properties of these new states, such as coherence and minimum-uncertainty, exhibit noticeable irregularities. This behaviour stems from the non-analyticity of the characteristic deformation function of the NLCS: indeed, such nonlinear deformation function of the number operator is non-analytic for the above values of $\eta$.\\
This point, which does not appear to have been tackled systematically so far, should be kept into
account also in view of possible experimental realizations. In fact, since the very structure of the Vogel NLCS and their properties are lost or compromised near these singular points, the interested experimentalist should fix the actual value of $\eta$ far from them. On the other hand, the theoretical predictions of this paper could be tested through direct observation of the deteriorated coherence property of the NLCS in a range of $\eta$ which includes one of the critical values.

The paper is organized as follows: the Hamiltonian of a trapped ion driven by two lasers is illustrated in Section 2, while in Section 3 it is shown how the $k$-boson NLCS are obtained resorting to standard tools of quantum optics, i.e., the interaction representation and the rotating wave approximation (RWA). Section 4 is devoted to the particular case of the single-boson NLCS, whose expansion in terms of Fock states is utilized in Section 5 to investigate the coherence and minimum-uncertainty properties in the singularity regions of the deformation function.

\section{The physical system}

The properties of the ion-trap systems and of their Hamiltonians in
various physical regimes have been investigated both theoretically
\cite{VOGEL1}-\cite{GER} and experimentally \cite{meek}-\cite{exp2}. As for the
construction of coherent states for ion-trap systems, it has been
proved that single-boson NLCS can be realized by a two-level ion
trapped in an external harmonic oscillator potential and interacting
with two laser fields \cite{MATOS1}.

Let $H$ $=$ $H_0$ $+$ $\displaystyle H_{\textrm{int}}$ be the Hamiltonian of the system. The free Hamiltonian $H_0$ describes the motion of both the internal electronic and external vibrational degrees of freedom of the ion, while the interaction Hamiltonian $H_{\textrm{int}}$ corresponds to the interaction of the ion with the electromagnetic fields of the driving lasers. The internal states of the ion are the ground state $|g\rangle$ and the excited state $|e\rangle$, whose transition frequency is $\omega$ $=$ $\omega_e - \omega_g$. For $\nu$ the frequency of the trap harmonic potential, $a$ the annihilation operator for the vibrational motion of the trapped ion, $\hat{n}$ $=$ $a^{\dagger} a$ and $\sigma_z$ $=$ $|e\rangle\langle e|$ - $|g\rangle\langle g|$ the Pauli matrix, which acts as a third component pseudospin operator, $H_0$ $=$ $\hbar \nu \hat{n} + \hbar \omega \sigma_z / 2$. Here the zero energy level of the ion has been set at halfway between $|g\rangle$ and $|e\rangle$. The interaction Hamiltonian reads instead
\begin{equation}
H_{\textrm{int}} \,=\, \lambda \left[E_0 \, e^{i(k_0 x - \omega_0 t)} + E_1 \, e^{i(k_1 x - \omega_1 t)} \right] S_+ + {\textrm{H.c.}} \, , \label{HAMINTER}
\end{equation}
where $\lambda$ is the dipole coupling matrix element, $E_0$, $E_1$, and $\omega_0$, $\omega_1$ are the amplitudes and the frequencies of the lasers fields, $k_0$, $k_1$ their wavectors. $x$ $=$ $\eta(a^{\dagger} + a)/k_L$ is the position operator of the ion center of mass, where $k_L$ $\approx$ $k_0 \approx k_1$ is the common value of both lasers wavevectors, $\eta = k_L \sqrt{\hbar/(2 M \nu)}$ is the Lamb-Dicke parameter, with $M$ the ion mass, and $S_+$ $=$ $(\sigma_x+i\sigma_y)/2$ $=$ $|e\rangle\langle g|$, $S_-$ $=$ $S_+^{\dagger}$ the two-level ion electronic flip operators. For the lasers tunings one sets $\omega_0 = \omega$, which corresponds to the resonant condition of the first laser, and $\omega_1 = \omega - k \nu$, which means that the second laser is tuned on $k$-th lower (or red-sideband) vibrational level, with $k$ an integer $\ge$ 1. From (\ref{HAMINTER}) one then obtains
\begin{equation}
H_{\textrm{int}} \,=\, \lambda e^{- i \omega t} \left[ E_0 \, e^{i\eta(a^{\dagger} + a)} + e^{i k \nu t} E_1 \, e^{i\eta(a^{\dagger} + a)} \right] S_+ + {\textrm{H.c.}} \; . \label{HAMINTAPPROX}
\end{equation}

\section{The $k$-boson NLCS}

The idea underlying the construction of nonlinear bosonic coherent states from the interaction Hamiltonian (\ref{HAMINTAPPROX}) can be summarized as follows:
\begin{enumerate}
    \item One constructs first the interaction representation of Hamiltonian (\ref{HAMINTAPPROX}), $\displaystyle \cal{H}_{\textrm{int}}$ $=$ $\displaystyle U^{\dagger} (t) H_{\textrm{int}} U(t)$ $\propto$ $\displaystyle (F_k \, S_+ + {\textrm{H.c.}})$, where $U(t)$ $=$ $\displaystyle \exp \left(- i H_0 t / \hbar \right)$ and $F_k$ is a non-Hermitian operator acting in the Fock space;
    \item resorting to the RWA $\displaystyle \cal{H}_{\textrm{int}}$ is then put in the form $\displaystyle {\cal{H}}_{\textrm{int}}^{(0)}$ $\propto$ $\displaystyle (F_k^{(0)} \, S_+ + {\textrm{H.c.}})$, the operator $F_k^{(0)}$ being the time-independent component of $F_k$;
    \item the zero-energy state $|\xi_k\rangle$ of ${\cal{H}}_{\textrm{int}}^{(0)}$, $\langle\xi_k|{\cal{H}}_{\textrm{int}}^{(0)}|\xi_k\rangle$ $=$ 0, is obtained from the condition $\displaystyle F_k^{(0)} |\xi_k\rangle$ $=$ 0;
    \item the zero-energy state $|\xi_k\rangle$ is recognized to be the eigenstate of the annihilation operator
\begin{equation}
A_k \,=\, f_k(\hat{n}) a^k \; , \label{AF}
\end{equation}
with the appropriate expression of the nonlinear function $f_k(\hat{n})$, i.e., the $|\xi_k\rangle$ are NLCS.
\end{enumerate}

Detailing now the above steps one obtains
\begin{equation}
{\cal{H}_{\textrm{int}}} \,=\, \lambda E_1 e^{-\eta^2/2} \left( F_k S_+ + {\textrm{H.c.}} \right) \; , \label{HAMINT2}
\end{equation}
with
\begin{equation}
F_k \,=\, \left(\frac{E_0}{E_1} + e^{i \nu k t} \right) \sum^{\infty}_{\ell,m=0} \frac{(i \eta)^{m+\ell}}{m! \ell!} e^{i \nu (m-\ell) t} {a^{\dagger}}^m a^\ell \; . \label{FK}
\end{equation}
Operator $F_k$ in (\ref{FK}) can be written in the form

\begin{equation}
F_k \,=\, \left(\frac{E_0}{E_1} + e^{i \nu k t} \right) \left[ \sum^{\infty}_{r=1} (i\eta)^r \left( e^{- i \nu r t} Z_r + {\textrm{H.c.}} \right) + L_{\hat{n}}^{(0)} (\eta^2) \right] \; , \label{FKZ}
\end{equation}
where $Z_r$ $=$ $\displaystyle \frac{L_{\hat{n}}^{(r)} (\eta^2)}{(\hat{n}+1) \cdots (\hat{n}+r)} a^r$ and $\displaystyle L_{\hat{n}}^{(m)} (\eta^2)$, $m$ integer $\ge$ 0, is the `Laguerre polynomial operator'. By this one means an appropriate integral representation\footnote{For example $\displaystyle L_{\hat{n}}^{(m)} (x)$ $=$ $\displaystyle \frac{1}{2 \pi i} \oint_{C} \frac{e^{-xz/(1-z)} \, z^{-\hat{n}}}{(1-z)^{m+1}} \frac{\textrm{d}z}{z}$, where $C$ is a closed contour centered at the origin of the complex plane, whose radius $|z|$ is $<$ 1.}, whose action in the Fock space results into the generalized Laguerre polynomial $\displaystyle L_n^{(m)} (\eta^2)$.

Applying the RWA to Hamiltonian (\ref{HAMINT2}) implies that all terms oscillating with frequency $ \geq \nu$ are disregarded in (\ref{FKZ}) so as to reduce $F_k$ to $F_k^{(0)}$ $=$ $\displaystyle \frac{E_0}{E_1} L_{\hat{n}}^{(0)} (\eta^2) + i \eta Z_k$, which leads to the explicit form of the RWA Hamiltonian
\begin{equation}
{\cal{H}}_{\textrm{int}}^{(0)} \,=\, \lambda E_1 e^{-\eta^2/2} \left( F_k^{(0)} \, S_+ + {\textrm{H.c.}} \right) \; . \label{INTGEN}
\end{equation}

States $|\xi_k\rangle$ are the coherent states defined as the eigenstates of the operator (\ref{AF})
\begin{equation}
A_k \, |\xi_k\rangle \,=\, k \, \xi_k \, |\xi_k\rangle \; , \label{EIGENAF}
\end{equation}
where $\xi_k$ $=$ $\displaystyle - (- i/\eta)^k (E_0/E_1)$ is the experimentally controllable parameter, while the nonlinear operator
\begin{equation}
f_k(\hat{n};\eta^2) \,=\, \prod^{k}_{j=1} (\hat{n}+j)^{-1} \, \frac{L_{\hat{n}}^{(k)} (\eta^2)}{L_{\hat{n}}^{(0)} (\eta^2)} \; . \label{FN}
\end{equation}
Manifestly, the function in (\ref{FN}), that for the sake of simplicity we shall henceforth denote simply as $f_k(\hat{n})$ as in (\ref{AF}), is non-analytic over the whole space of the parameter $\eta$.

With $\displaystyle |\xi_k\rangle$ $=$ $\displaystyle \sum^{\infty}_{n=0} c_n^{(k)} |n\rangle$, the condition $\displaystyle F_k^{(0)} |\xi_k\rangle$ $=$ 0 leads to $\displaystyle  |\xi_k\rangle$ $=$ $\displaystyle \sum^{k-1}_{\ell=0} |\xi_k\rangle_\ell$, where the states $|\xi_k\rangle_\ell$ $=$ $\displaystyle \sum^{\infty}_{n=0} c_{nk+\ell}^{(k)} |nk+\ell\rangle$ are fixed by the corresponding sets of coefficients
\begin{equation}
c_{nk+\ell}^{(k)} \,=\, \frac{\sqrt{\ell!} \, \xi_k^n}{\sqrt{(nk+\ell)!}} \prod^{n-1}_{m=0} [f_k(\ell+mk)]^{-1} c_\ell^{(k)} \; . \label{CNKL}
\end{equation}
The normalization constants $\displaystyle c_\ell^{(k)}$ chosen to be real in order to retrieve the Glauber states when $k = 1$ and $f \equiv 1$, are
\begin{equation}
c_{\ell}^{(k)} \,=\, \left\{ \sum^{\infty}_{n=0} \frac{\ell! \, |\xi_k|^{2n}}{(nk+\ell)!} \prod^{n-1}_{m=0} [f_k(\ell+mk)]^{-2} \right\}^{- \frac{1}{2}} \; . \label{NORMKL}
\end{equation}
In (\ref{CNKL}), (\ref{NORMKL}) the scalar functions $f_k(\ell+mk)$ are obtained from the action of operator (\ref{FN}) in the Fock space, and the convention $\displaystyle \prod^{n-1}_{m=0} \doteq 1$ for $n = 0$ is adopted.

It is readily verified that also the states $|\xi_k\rangle_\ell$ are eigenstates of operator $A_k$ defined in (\ref{AF}), $\displaystyle A_k |\xi_k\rangle_\ell$ $=$ $\displaystyle \xi_k |\xi_k\rangle_\ell$, with the $k$-degenerate eigenvalue $\xi_k$. Since $\displaystyle _j\langle\xi_k|\xi_k\rangle_\ell$ $=$ $\delta_{j \ell}$, they form an orthonormal basis of a $k$-dimensional Hilbert space.

\section{The single-boson NLCS}
In the nondegenerate case $k = 1$ the construction reported in \cite{MATOS1} is restored. The notation adopted so far can be simplified by dropping the index $k$ throughout.

\subsection{The states $|\xi\rangle$ as eigenstates of $A$ $=$ $f(\hat{n}) a$}

In view of (\ref{EIGENAF}) $|\xi\rangle$ is by construction the eigenstate of the operator $A$ $=$ $f(\hat{n}) a$, the eigenvalue being $\xi$ $=$ $\displaystyle i E_0/(\eta E_1)$. The relevant commutation relations are $\displaystyle \left[\hat{n},A\right]$ $=$ $- A$, $\displaystyle \left[\hat{n},A^{\dagger}\right]$ $=$ $\displaystyle A^{\dagger}$ and
\begin{equation}
\left[A,A^{\dagger}\right] \,=\, (\hat{n}+1) f^2(\hat{n}) - \hat{n} f^2(\hat{n}-1) \doteq C(\hat{n}) \; . \label{COM_AA}
\end{equation}
In terms of Fock states
\begin{equation}
|\xi\rangle \,=\, {\cal{N}} \sum^{\infty}_{n=0} c_n |n\rangle \;,\; {\cal{N}} \,=\, \left[ \sum^{\infty}_{n=0} \left| c_{n} \right|^2 \right]^{-\frac{1}{2}}
\; , \label{COHER}
\end{equation}
where the normalization constant ${\cal{N}}$ is but $c_{0}^{(1)}$ in Eq. (\ref{NORMKL}) and (cfr. (\ref{CNKL}))
\begin{equation}
c_n \,=\, \frac{\xi^n}{\sqrt{n!}} \prod^{n-1}_{j=0} \left[ f(j) \right]^{-1} \,=\, \xi^n \sqrt{n!} \prod^{n-1}_{j=0} \frac{L_{j}^{(0)} (\eta^2)}{L_{j}^{(1)} (\eta^2)} \; , \label{GAMMA}
\end{equation}
with $f(j)$ $=$ $\displaystyle L_{j}^{(1)} (\eta^2) / [(j+1) \, L_{j}^{(0)} (\eta^2)]$ and $c_0$ $\doteq$ 1. For $m \ge 1$, the following recurrence formula holds for the coefficients (\ref{GAMMA})
\begin{equation}
c_{n + m} \,=\, \xi^m \sqrt{\frac{n!}{(n+m)!}} \prod^{m-1}_{\ell=0} [f(n+\ell)]^{-1} c_n \; . \label{GAMRECURR}
\end{equation}

Notice that for $\eta^2 = 0$, $|\xi\rangle$ reduce to the Glauber states as in this case $f(j) = 1, \, \forall \, j$.

\subsection{The states $|\xi\rangle$ as minimum-uncertainty states}

For dimensionless Hermitian operators $\displaystyle Q$ $=$ $\displaystyle (A^{\dagger} + A)/\sqrt{2}$, $\displaystyle P$ $=$ $\displaystyle i (A^{\dagger} - A)/\sqrt{2}$, in view of (\ref{COM_AA}), one has $\displaystyle [Q,P]$ $=$ $i C(\hat{n})$. Of course $\displaystyle Q$ and $\displaystyle P$ are not the usual phase space operators, i.e., they are not the center of mass position and momentum of the ion, but they are introduced here as a convenient device to test the minimum-uncertainty property of states $|\xi\rangle$. Indeed, calculating the variances $\displaystyle (\Delta Q)_{\xi}^2$, $\displaystyle (\Delta P)_{\xi}^2$ in states $|\xi\rangle$, one finds the Heisenberg uncertainty relation $\displaystyle (\Delta Q)^2 (\Delta P)^2$ $\ge$ $\displaystyle |\langle [Q, P] \rangle|^2 /4$ with an equal sign, as
\begin{equation}
(\Delta Q)_{\xi}^2 \,=\, (\Delta P)_{\xi}^2 \,=\, \frac{1}{2} \left| \langle\xi| C(\hat{n}) |\xi\rangle \right| \; . \label{XPINTELL}
\end{equation}
Here $\langle\xi| C(\hat{n}) |\xi\rangle$ $=$ $\displaystyle {\cal{N}}^2 \sum^{\infty}_{n=0} \left| c_{n} \right|^2 C(n)$ $=$ $\displaystyle \sum^{\infty}_{n=0} p_n C(n)$, where $p_n$ $=$ $|c_{n}|^2 {\cal{N}}^2$, $0\le p_n\le 1$ $\forall \, n$, and $C(n)$ $=$ $(n+1) f^2(n) - n f^2(n-1)$. Eq. (\ref{XPINTELL}) confirms that states (\ref{COHER}) are intelligent states. On the other hand, they are not generalized coherent states since, in view of (\ref{COM_AA}), $A$ and $A^{\dagger}$ do not generate a finite-dimensional algebra.

\section{The singularities}

Eq. (\ref{GAMMA}) shows that, due to the presence of non-analytic
functions $f(j)$, the coefficients $c_n$ possess $n(n-1)/2$ zeros
and as many poles corresponding to the roots of the $n$ Laguerre
polynomials $\displaystyle L_{j}^{(0)} (\eta^2)$ and $\displaystyle
L_{j}^{(1)} (\eta^2)$, respectively. One expects that the structure
of states (\ref{COHER}) is strongly modified if the values of
$\eta^2$ result into a vanishing or singular $c_n$.

The problem, referred to briefly in \cite{MANKO}, is addressed in
\cite{MATOS3} in relation to the representation of the Fock states
in terms of single-boson NLCS.

\subsection{Effect of a single zero of $c_n$}

By `single zero' of a given $c_n$, say $c_\nu$, one means that $\eta^2$ is such that $\displaystyle L_{\nu-1}^{(0)} (\eta^2)$ $=$ 0 and, equivalently, $\displaystyle f(\nu-1)$ $\rightarrow$ $\infty$. Since, from (\ref{GAMRECURR}), $\displaystyle c_{\nu+m}$ $=$ 0, $\forall$ $m \ge 1$, the sums in $|\xi\rangle$ and ${\cal{N}}$ can be truncated so that the states read
\begin{equation}
|\xi\rangle_{\nu} \,=\, {\cal{N}}_{\nu} \sum^{\nu-1}_{n=0} c_{n} |n\rangle \;,\; {\cal{N}}_{\nu} \,=\, \left[ \sum^{\nu-1}_{n=0} \left| c_{n} \right|^2 \right]^{-\frac{1}{2}} \; . \label{COHZERO}
\end{equation}

The vanishing of coefficient $\displaystyle c_{\nu}$ breaks the coherence of the NLCS $|\xi\rangle$ as the states (\ref{COHZERO}) do not satisfy any longer the eigenvalue equation of the annihilation operator. Indeed, $A \, |\xi\rangle_{\nu}$ $=$ $\displaystyle \xi \left( |\xi\rangle_{\nu} - {\cal{N}}_{\nu} c_{\nu-1} |\nu-1\rangle \right)$, where the rhs is different from $\displaystyle \xi |\xi\rangle_{\nu}$.

\begin{figure}[htb]
    \centering
        \includegraphics[width=0.85\textwidth]{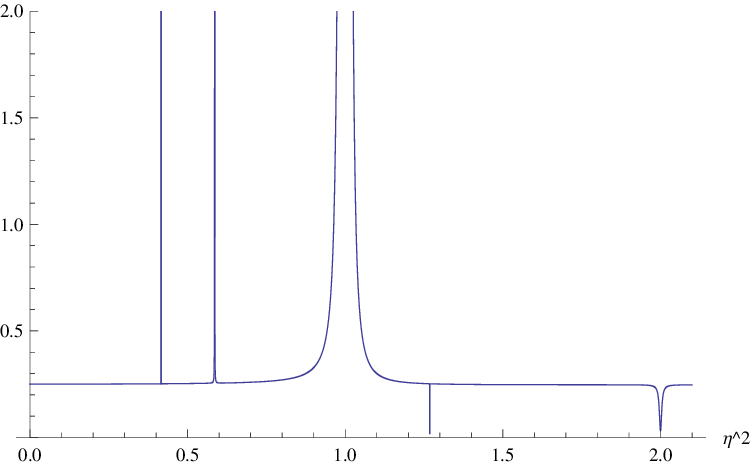}
    \caption{$(\Delta Q)_{\nu}^2$ $(\Delta R)_{\nu}^2$ vs. $\eta^2$ for $\nu = 4$, $|\xi| = 0.05$}
    \label{fig:heiszerozero5}
\end{figure}

\begin{figure}[htb]
    \centering
        \includegraphics[width=0.85\textwidth]{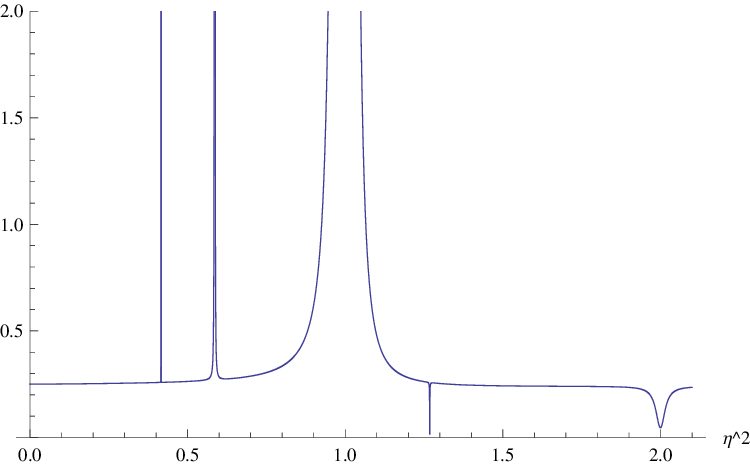}
    \caption{$(\Delta Q)_{\nu}^2$ $(\Delta R)_{\nu}^2$ vs. $\eta^2$ for $\nu = 4$, $|\xi| = 0.1$}
    \label{fig:heiszero1}
\end{figure}

As for the minimum-uncertainty property, the variances of operators $Q$ and $P$ in states (\ref{COHZERO}), $(\Delta Q)_{\nu}^2$  and $(\Delta P)_{\nu}^2$, differ markedly from expression (\ref{XPINTELL}). In fact, one has $\displaystyle (\Delta Q)_{\nu}^2$ $=$ $\Xi + \Lambda$, $\displaystyle (\Delta P)_{\nu}^2$ $=$ $\Xi - \Lambda$, where
\begin{eqnarray}
\Xi &=& |\xi|^2 p_{\nu-1}(\nu) \left[1 - p_{\nu-1}(\nu) \right] + \frac{1}{2} \sum^{\nu-1}_{n=0} p_n(\nu) C(n) \; , \nonumber\\
\Lambda &=& \frac{1}{2} \left( \xi^2 + {\bar{\xi}}^{ 2} \right) \left[p_{\nu-1}(\nu) - p_{\nu-1}^2(\nu) - p_{\nu-2}(\nu) \right] \; , \nonumber\
\end{eqnarray}
with $p_n(\nu)$ $=$ $\displaystyle |c_{n}|^2 {\cal{N}}_{\nu}^2$, $0 \le p_n(\nu) \le 1$ $\forall \, n, \nu$. The bar denotes complex conjugation. The product $(\Delta Q)_{\nu}^2$ $(\Delta P)_{\nu}^2$ $=$ $\displaystyle \Xi^2 - \Lambda^2$ has to be evaluated numerically. In Figs. \ref{fig:heiszerozero5} and \ref{fig:heiszero1} it is exemplified, in the range $0 < \eta^2 \le 2$, the case $c_4 = 0$, i.e., $\eta^2$ one of the roots of $\displaystyle L_3^{(0)} (\eta^2)$, considering $|\xi| = 0.05$ and $|\xi| = 0.1$, respectively. In fact, due to the product form of $c_n$ in (\ref{GAMMA}), the product of the variances diverges when $\eta^2$ is one of the roots of the Laguerre polynomials $L_j^{(0)} (\eta^2)$, $j = 1, 2, 3$, which lie in the considered range.

The results are significantly influenced by the parameter $\xi$. Indeed, Fig. \ref{fig:heiszerozero5} shows that, for a relatively small value of $\xi$, the regions where the product of the variances $\displaystyle (\Delta Q)_4^2 (\Delta P)_4^2$ diverges are very localized. Furthermore, the `classical' minimum-uncertainty value 1/4 is maintained over a wide range of values of $\eta^2$: specifically, in these regions the numerical calculations give $(\Delta Q)_{\nu}^2$ $=$ $(\Delta P)_{\nu}^2$ $\approx$ 1/2 and $C(\hat{n})$ $\approx$ $\mathbb{I}$. Therefore, the corresponding states (\ref{COHZERO}) are minimum-uncertainty states and can be conveniently utilized for QT purposes, while $Q$ and $P$ can be considered, approximately, canonically conjugate operators. On the other side, Fig. \ref{fig:heiszero1} and further numerical calculations prove that the minimum-uncertainty property is verified to a lesser degree for larger values of $\xi$. Numerical investigation shows as well that, as expected, the minimum-uncertainty feature depends also on the parameter $\nu$, in the sense that the lower the value of $\nu$, the wider the ranges where the minimum-uncertainty property is verified by the truncated states (\ref{COHZERO}).

When $\eta^2$ is one of the roots of the polynomials $L_j^{(1)} (\eta^2)$, $j = 1, 2, 3$, a different effect is registered. Figs. \ref{fig:heiszerozero5} and \ref{fig:heiszero1} show that the
variance product does not diverge and never reaches zero.

\subsection{Effect of a single pole of $c_n$}

By `single pole' of a given $c_n$, say $c_{\mu}$, one means that $\eta^2$ is such that $\displaystyle L_{\mu-1}^{(1)} (\eta^2)$ $=$ 0 and, equivalently, $\displaystyle f(\mu-1)$ $=$ 0. From the recurrence relation (\ref{GAMRECURR}) it follows that the coefficients $\displaystyle c_{\mu+m}$ $\rightarrow$ $\infty$ as rapidly as $c_{\mu}$, $\forall$ $m \ge 1$. The above conditions imply that one can disregard the first $\mu$ terms of the sums in $|\xi\rangle$ and ${\cal{N}}$ obtaining the following expressions of the states and their normalization constant
\begin{equation}
|\xi\rangle_{\mu} \,=\, {\cal{N}}_{\mu} \, \sum^{\infty}_{j=0} c_{\mu+j} |\mu+j\rangle \;,\, {\cal{N}}_{\mu} \,=\, \left[ \sum^{\infty}_{j=0} |c_{\mu+j}|^2 \right]^{-\frac{1}{2}} \; . \label{COHPOLO}
\end{equation}
Factorizing the singular terms $\displaystyle c_{\mu}$ and $\displaystyle |c_{\mu}|$ in (\ref{COHPOLO}) and resorting to (\ref{GAMRECURR}) one finds
$$
|\xi\rangle_{\mu} \,=\, i^\mu S_\mu^{-\frac{1}{2}} \, \sum^{\infty}_{j=0} \xi^j \sqrt{\frac{\mu!}{(\mu+j)!}} \prod^{j-1}_{\ell=0} [f(\mu+\ell)]^{-1} |\mu+j\rangle \; ,
$$
where $i^\mu = c_{\mu}/|c_{\mu}|$ is a global phase factor and
\begin{equation}
S_\mu \,=\, (|c_{\mu}|^2 {\cal{N}}^2_{\mu})^{-1} \,=\, \sum^{\infty}_{j=0} |\xi|^{2j} \frac{\mu!}{(\mu+j)!} \prod^{j-1}_{\ell=0}
[f(\mu+\ell)]^{-2} \; . \label{SERIES}
\end{equation}
The question whether or not $S_\mu$ converges, and hence the states (\ref{COHPOLO}) exist (in the sense that they are normalizable), can be tackled, for example, through the ratio test. According to this criterion the convergence of the series (\ref{SERIES}), written as $S_\mu$ $=$ $\displaystyle \sum^{\infty}_{j=0} s_j^{(\mu)}$, implies the evaluation of the ratio
$$
\frac{s_{j+1}^{(\mu)}}{s_j^{(\mu)}} \,=\, \frac{|\xi|^2}{(\mu+j+1) f ^2 (\mu+j)} \,=\, |\xi|^2 (\mu+j+1) \left[\frac{L_{\mu+j}^{(0)} (\eta^2) }{L_{\mu+j}^{(1)} (\eta^2)} \right]^2 \; ,
$$
when $j \rightarrow \infty$. Resorting to the asymptotic form for large degree Laguerre polynomials \cite{SZEGO}
$$
L_{n}^{(m)} (\eta^2) \, \approx \, \frac{e^{\eta^2/2}}{\sqrt{\pi}} \frac{n^{(m-1/2)/2}}{\eta^{(m+1/2)}} \cos \left(2 \eta \sqrt{n}-m \frac{\pi}{2}-\frac{\pi}{4}\right) \quad\quad n \gg 1 \; ,
$$
one can check that, upon setting $\alpha$ $=$ $\displaystyle \left(\frac{\pi}{8 \eta}\right)^2$,
\begin{itemize}
\item[\textit{i})]  $S_\mu$ diverges if $\alpha \in \mathbb{Q}$;
\item[\textit{ii})] if $\alpha$ is sufficiently irrational, i.e., $\displaystyle \left|\alpha-\frac{p}{q}\right| > \frac{\gamma}{q^\beta}$, for $p$, $q$ arbitrarily large integers, whose ratio best approximates $\alpha$, $\gamma \in \mathbb{R}$ and $\beta \gg 1$, then $S_\mu$ converges. For example, for $\beta = 1$, this happens provided $|\xi|^2 < 1$, i.e., $\eta > E_0/E_1$.
\end{itemize}
Since $\eta$ is a physical parameter of the system, whose square is here assumed to be a root of $\displaystyle L_{\mu-1}^{(1)} (x)$, it is unlikely that $\alpha$ is rational, therefore case $ii)$ is what one should expect.

Writing the action of $A$ on the states (\ref{COHPOLO}) one has, in the single pole case,
\begin{equation}
A \, |\xi\rangle_{\mu} \,=\, \xi \left( |\xi\rangle_{\mu} + {\cal{N}}_{\mu} c_{\mu-1} |\mu-1\rangle \right) \; , \label{APOLO}
\end{equation}
while the minimum-uncertainty property in states (\ref{COHPOLO}) reads
\begin{equation}
(\Delta Q)_{\mu}^2 \,=\, (\Delta P)_{\mu}^2 \,=\, |\xi|^2 |c_{\mu-1}|^2 {\cal{N}}_{\mu}^2 + \frac{1}{2} \sum^{\infty}_{n=\mu} p_n(\mu) C(n) \; , \label{HEISPOLO}
\end{equation}
where $p_n(\mu)$ $=$ $\displaystyle |c_{n}|^2 {\cal{N}}_{\mu}^2$, $0 \le p_n(\mu) \le 1$ $\forall \, n, \mu$. Notice that, since $\displaystyle {\cal{N}}_{\mu}$ is vanishingly small, from (\ref{APOLO}) and (\ref{HEISPOLO}) it follows that the coherence and the minimum-uncertainty properties are not fully compromised when a single pole is encountered. This result is physically plausible as states (\ref{COHPOLO}) are still an infinite superposition of Fock states and one retrieves, in the context of the NLCS, the property of the Glauber states of being eigenstates of any power of the annihilation operator $a$.

\section{Concluding remarks}
In this work the properties of the NLCS constructed for a two-level laser-driven trapped ion have been investigated in the framework of the usual quantum optics approach based on both the interaction picture and the RWA. Due to the manifest non-analyticity of the function in the deformed annihilation operator, one expects that the fundamental quantum features of these states can exhibit significant differences with respect to the customary coherent states. The single-boson case has been analyzed since the $k$-boson NLCS are expected (see Eqs. (\ref{FN})-(\ref{NORMKL}) ) to exhibit exactly the same problems. Specifically:
\begin{itemize}
    \item if the squared Lamb-Dicke parameter happens to be a root of a given Laguerre polynomial $\displaystyle L_{\nu-1}^{(0)} (\eta^2)$, then states (\ref{COHER}) reduce to the finite sum of Fock states (\ref{COHZERO}) and the coherence and the minimum-uncertainty properties are compromised. However, the latter feature still holds over a wide range of typical values of $\eta$ for relatively small values of $\xi$ and if $\nu$ is not exceedingly high (i.e., one does not consider highly excited vibrational levels);
    \item if the squared Lamb-Dicke parameter happens to be a root of a given Laguerre polynomial $\displaystyle L_{\mu-1}^{(1)} (\eta^2)$, then states (\ref{COHER}) transform into the infinite sum of Fock states (\ref{COHPOLO}). The coherence and minimum-uncertainty properties are retained when the series (\ref{SERIES}) converges, which was shown to be the case one should expect.
\end{itemize}   
   
To summarize, the conclusions of the present analysis should be kept into consideration when implementing the ion trap NLCS. Indeed, such states have already been realized experimentally \cite{meek} and can find prospective applications ranging from quantum information to basic science. It is therefore important to bear in mind that their very structure and their specific properties can be lost or compromised if the physical system is too close to critical values of the Lamb-Dicke parameter induced by the zeros of the relevant Laguerre polynomials. This can happen also on account of the finite accuracy in fixing the control parameter $\eta$ when preparing single and \textit{k}-boson NLCS for actual experiments.

\section{Acknowledgments}
Two of the authors (M. R. and M. G.) would like to acknowledge the financial support of Regione Piemonte through the Research Project E14.



\bibliographystyle{model1a-num-names}
\bibliography{<your-bib-database>}







\end{document}